\documentclass[twocolumn, nofootinbib]{revtex4}

\usepackage{url}
\usepackage{epsfig}
\usepackage{amsfonts}
\usepackage{graphicx}
\usepackage{epsfig}
\usepackage{eepic}
\usepackage{amsmath}
\usepackage{amssymb}
\usepackage{color}
\usepackage{bbm}
\usepackage{dcolumn}
\usepackage{bm}
\usepackage{ulem}
\usepackage{mathrsfs}
\usepackage{bbold}
\usepackage{datetime}

\usepackage{overpic} 
\usepackage{rotating}
\usepackage[usenames,dvipsnames]{xcolor}
\usepackage[colorlinks=true,citecolor=Magenta,linkcolor=Green,urlcolor=Blue]{hyperref}
\usepackage{lipsum} 
\usepackage{etoolbox} 
\usepackage[capitalize]{cleveref}

\usepackage[font=small,labelfont=bf,justification=raggedright]{caption}

\begin{document}

\title {Holographic Abrikosov lattice: vortex matter from black hole}

\author{Chuan-Yin Xia$^1$}\email{chuanyinxia@foxmail.com}
\author{Hua-Bi Zeng$^1$}\email{hbzeng@yzu.edu.cn}
\author{Yu Tian$^{2,3}$}\email{ytian@ucas.ac.cn}
\author{Chiang-Mei Chen$^4$}\email{cmchen@phy.ncu.edu.tw}
\author{Jan Zaanen$^5$}\email{jan@lorentz.leidenuniv.nl}

\affiliation{$^1$Center for Gravitation and Cosmology, College of Physical Science and Technology, Yangzhou University, Yangzhou 225009, China}
\affiliation{$^2$School of Physics, University of Chinese Academy of Sciences, Beijing 100049, China}
\affiliation{$^3$Institute of Theoretical Physics, Chinese Academy of Sciences, Beijing 100190, China}
\affiliation{$^4$Department of Physics, Center for High Energy and High Field Physics (CHiP), National Central University, Chungli 32001, Taiwan}
\affiliation{$^5$Institute Lorentz for Theoretical Physics, Leiden University, Leiden, The Netherlands}

\begin{abstract}
The AdS/CFT correspondence provides a unique way to study the vortex matter phases in superconductors. We solved the nonlinear equations of motion for the Abelain-Higgs theory living on the AdS$_4$ black hole boundary that is dual to a two dimensional strongly coupled type II superconductor at temperature $T$ with a perpendicular external uniform magnetic field $B_0$. We found the associated two critical magnetic fields, $B_{c1}(T)$ and $B_{c2}(T)$. For $B_0 < B_{c1}(T)$ the magnetic field will be expelled out by the superconductor resembling the Meissner effect and the superconductivity will be destroyed when $B_0 > B_{c2}(T)$. The Abrikosov lattice appears in the range $B_{c1}(T) < B_0 < B_{c2}(T)$ including, due to the finite size and boundary effect, several kinds of configurations such as hexagonal, square and slightly irregular square lattices, when the magnetic field is increased. The upper and lower critical fields behave as inverse squares of coherence length and magnetic penetration depth respectively which matches the well known consensus.
\end{abstract}


\maketitle

\textit{Introduction} ---
A well-known property of the type II superconductors is the quantization of the magnetic flux in the mixed state, where the magnetic field penetrates the sample as vortices forming an Abrikosov lattice in which each vortex carrying one thread of magnetic field with the flux $\Phi_0 = h c/2 e$~\cite{Abrikosov}. Close to the transition temperature, the formation of vortex lattice can be simulated by the time dependent Ginzburg-Landau (GL) equation with parameter $\kappa > 1/\sqrt{2}$. The GL parameter $\kappa$ is defined as the ratio of the magnetic penetration depth $\lambda$ to the coherence length $\xi$ of the order parameter, $\kappa = \lambda/\xi$, and it can be calculated from the microscopic parameters of the material within the Bardeen–Cooper–Schrieffer (BCS) theory. In a type I superconductor, $\kappa < 1/\sqrt{2}$, the interaction between vortices is purely attractive which results in their fusion into macroscopic normal domains in the intermediate state, then no stable vortices appear. On the contrary in a type II superconductor, $\kappa > 1/\sqrt{2}$, the interaction between vortices is purely repulsive so the vortices are stable and form a lattice in the mixed state~\cite{Brandt, Bogomolnyi1, Bogomolnyi2, Jacobs, Beekman:2011tn}. For review of GL theory of type II superconductors under magnetic field, please refer to~\cite{Ginzburg, Gennes, Saint, Blatter:1994zz, Tinkham, Rosenstein}.

In the AdS/CFT correspondence framework~\cite{Maldacena, Gubser, Witten}, the holographic version of the superconducting model was firstly proposed in~\cite{Gubser2008, Hartnoll}. The holographic superconductor model includes a charged scalar field living in an AdS planar black hole. The scalar field, when the temperature of the black hole is lower enough, can have a nonzero profile with lower free energy than the trivial zero solution. Such an $U(1)$ symmetry broken configuration is dual to a superconducting state in the boundary field theory by computing the conductivity via the AdS/CFT correspondence dictionary~\cite{Hartnoll}. The computation of the associated GL parameter $\kappa$ indicates that such a superconductor is always of type II in the probe limit~\cite{Umeh:2009ea, Hartnoll:2008kx, huabi}. Afterward many efforts have been devoted to consider external magnetic field effects~\cite{Nakano:2008xc, Albash:2008eh, Albash:2009ix}, in particular, to find the stable vortex state in the $s$-wave holographic superconductor, by solving the time independent equation of motion (EoMs) for the scalar and gauge fields in the bulk spacetime~\cite{Montull, Domenech, Keranen, Albash:2009iq, Maeda, Aristomenis}. Due to the highly nonlinear properties of the EoMs, single vortex solution, rather than vortex lattice, had been obtained~\cite{Montull, Domenech, Keranen, Albash:2009iq}, while a static vortex lattice solution was obtained by the perturbative method~\cite{Maeda, Aristomenis}.

Rather than solving the time independent EoMs, the vortex lattice states can also be obtained by studying the dynamics of a homogeneous superconductor in the presence of a uniform magnetic field by solving the full time dependent EoMs of the holographic superconductor model. The equilibrium phase within fixed temperature and magnetic field can be obtained by finding the final stable configuration which does not change in time anymore. This is very similar to the vortex lattice formation dynamics simulation by solving the time dependent Gindzburg-Landau equation~\cite{Dorsey, Du1}. In this letter, we report the results of the equilibrium vortex lattice phases, the magnetization curve and two critical fields in the phase diagram. We also address temperature dependence of the magnetic penetration depth, the coherence length and two critical fields which match the experimental observations.

\textit{Holographic Model} ---
The action of the Abelian-Higgs model in AdS$_4$ black holes, in the unit $\hbar = c = G_N = 1$, reads~\cite{Gubser2008, Hartnoll}
\begin{equation}
S = \int d^4x \sqrt{-g} \left( - \frac{1}{4} F^2 - |D \Psi|^2 - m^2 |\Psi|^2 \right),
\end{equation}
where $F_{\mu\nu} = \partial_\mu A_\nu - \partial_\nu A_\mu$ and $D_\mu = \nabla_\mu - i q A_\mu$ with the charge of scalar field, as a Cooper pair, $q = 2 e$. The AdS$_4$ black hole background in the Eddington-Finkelstein coordinates is
\begin{equation}
ds^2 = \frac{\ell^2}{z^2} \left( - f(z) dt^2 - 2 dt dz + dx^2 + dy^2 \right),
\end{equation}
in which $\ell$ is the AdS radius, $z$ is the radial coordinate of the AdS bulk and $f(z) = 1 - (z/z_h)^3$. Thus, $z = 0$ is the AdS boundary while $z = z_h$ is the horizon. The dual field theory lives at $z = 0$, and the information needed for the dual superconductor can be read from the behaviors of the fields on the boundary by solving the dynamic coupled equations of motion for $\Psi$ and $A_\mu$
\begin{equation}
(D^2 - m^2) \Psi = 0, \quad \nabla_\mu F^{\mu\nu} = i \Psi^* D^\nu \Psi - i \Psi (D^\nu \Psi)^*.
\end{equation}
Implicitly, the spontaneous broken of the local $U(1)$ symmetry for the field theory is induced by a nonzero expectation value $\Psi^{(2)}$ of the scalar operator dual to $\Psi$ in the bulk, which is read from the asymptotic behavior of $\Psi$ on the boundary
\begin{equation}
\Psi(z \sim 0, t, x, y) \approx \Psi^{(1)}(t, x, y) \, z + \Psi^{(2)}(t, x, y) \, z^2,
\end{equation}
where the source $\Psi^{(1)}$ is set to be zero as a boundary condition when solving the model. Furthermore, in order to introduce a magnetic field in the dual holographic superconductor, the gauge fields on the boundary should be dynamic. With the gauge fixing $A_z = 0$, the behavior of the gauge fields on the boundary is
\begin{equation}
A_\mu(z \sim 0, t, x, y) \approx a_\mu(t, x, y) + b_\mu(t, x, y) \, z,
\end{equation}
in which $a_\mu$ can be regarded as the gauge field of the boundary theory, while $b_\mu$ is related to the current $j_\mu$ as $j_\mu = - b_\mu - \partial_\mu a_t + \partial_t a_\mu$ in the Eddington coordinate following the holographic dictionary. We control the charge density $\rho = - b_t$ in equivalent to turn the temperature. In the superconductor case, we fix $j_x = j_y = 0$ as the Neumann boundary condition for $A_x$ and $A_y$ at $z = 0$. Instead, in the superfluid case, the Dirichlet boundary condition $a_x = a_y = 0$ was imposed and the vortex lattice solution has been found in~\cite{Dias, zeng, XL, Wittmer, Ewerz, Ankur}.

Similar to the experimental setup for generating vortices, we prepare a homogeneous superconducting state as the initial configuration, and then an uniform external magnetic field is applied suddenly to the sample at $t = 0$ by turning on $A_x(t = 0, z, x, y) = - B_0 y/2$ and $A_y(t = 0, z, x, y) = B_0 x/2$. We firstly prepare the initial homogeneous superconducting state at a fixed temperature by the Newton--Raphson method. Its evolution under an external magnetic field can be simulated by combining a Runge-Kutta method in the time direction and a Chebyshev spectral method for the other three coordinates $z, x, y$, similar to previous work on vortex lattice formation in a rotating holographic superfluid~\cite{zeng}.

\begin{figure}
\centering
\includegraphics[trim=5.5cm 10.8cm 5cm 11cm, clip=true, scale=0.8, angle=0]{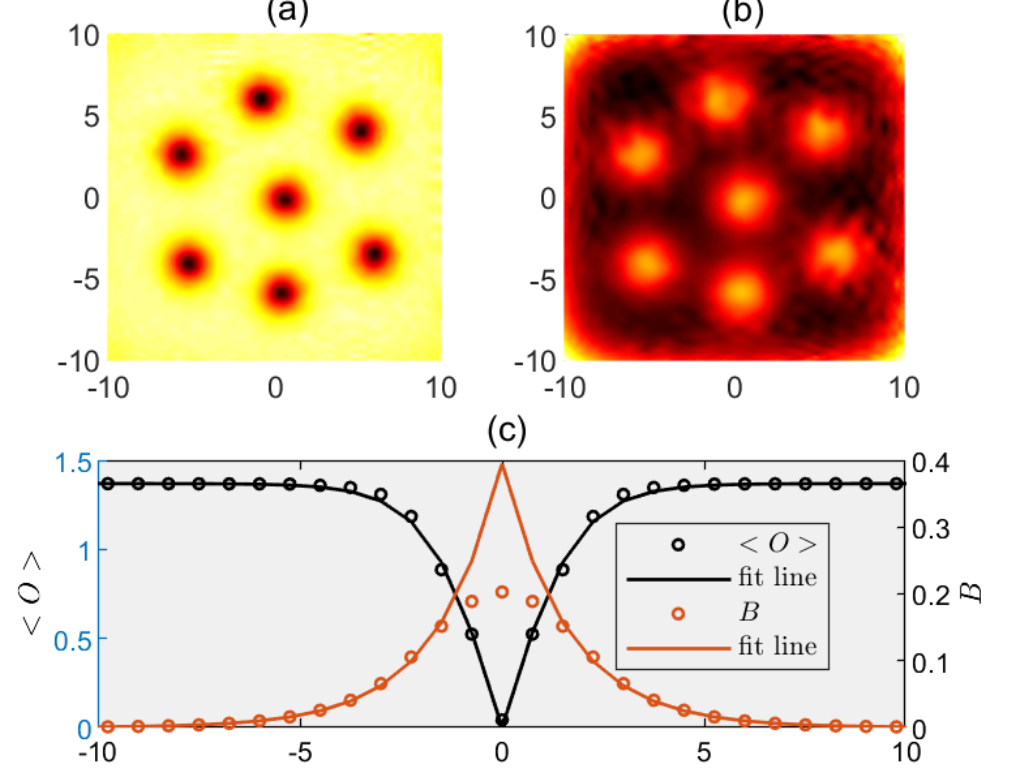}
\caption{A hexagonal lattice at $T = 0.95 \, T_c, \; B_0 = 0.428$: configurations of (a) the order parameter, (b) the magnetic field, and (c) the radial profiles of the order parameter and magnetic field in a single vortex.
} \label{fig_config1}
\end{figure}

\textit{Abrikosov Lattices} ---
For the numerical simulation, we chose $\ell = 1, m^2 = -2$ and $q = 1$ (implying $\Phi_0 = 2\pi$). In Fig.~\ref{fig_config1} we present a typical hexagonal Abrikosov lattice solution as a final stable state does not change any more for a sufficient long time in the dynamic process, when the temperature is very close to $T_c$, and the configuration of order parameter and magnetic field for one single vortex. Widths of the flux lines $\lambda$ and of the order parameter defects $\xi$ can be fitted from the profile of magnetic field $B(r) \sim 0.3949 \exp(-r/\lambda)$ and the expectation value of the order parameter $\langle O(r) \rangle \sim 1.3693 \tanh(r/\sqrt2 \xi)$. From Fig.~\ref{fig_config1}, we can estimate the values $\lambda \sim 1.579$ and $\xi \sim 1.1$, respectively. Thus, the GL parameter is $\kappa \sim 1.435$, which belongs to type II superconductors.

According to the GL theory analysis, the lattices with equilateral triangles admit a slightly lower free energy than the square ones. It is interesting that this result agrees with that of a simple argument based on the fact that the triangular array is a ``closed-packed'' one, in which each vortex is surrounded by a hexagonal array of other vortices. In this array, the nearest neighbor distance can be evaluated from the averaged value of the magnetic field in a vortex $\langle B \rangle$ as
\begin{equation} \label{atrig}
a_{\triangle} = \left( \frac{4}{3} \right)^{\frac14} a_{\square} \approx \left( \frac{4}{3} \right)^{\frac14} \left( \frac{\Phi_0}{\langle B \rangle} \right)^{\frac12}.
\end{equation}
Thus, for a given flux density, $a_{\square} < a_{\triangle}$. Taking into account the mutual repulsion of the vortices, it is reasonable that the structure with the greatest separation of the nearest neighbors would be favored. From Eq.~(\ref{atrig}) the distance between two nearest vortices can be computed as $a_{\triangle} \approx 1.075 \sqrt{2 \pi/0.178} \approx 6.39$, 
close to the numerical simulation $a_\triangle \approx 6.63$.

\begin{figure}
\centering
\includegraphics[trim=3.3cm 8cm 3cm 7.5cm, clip=true, scale=0.5, angle=0]{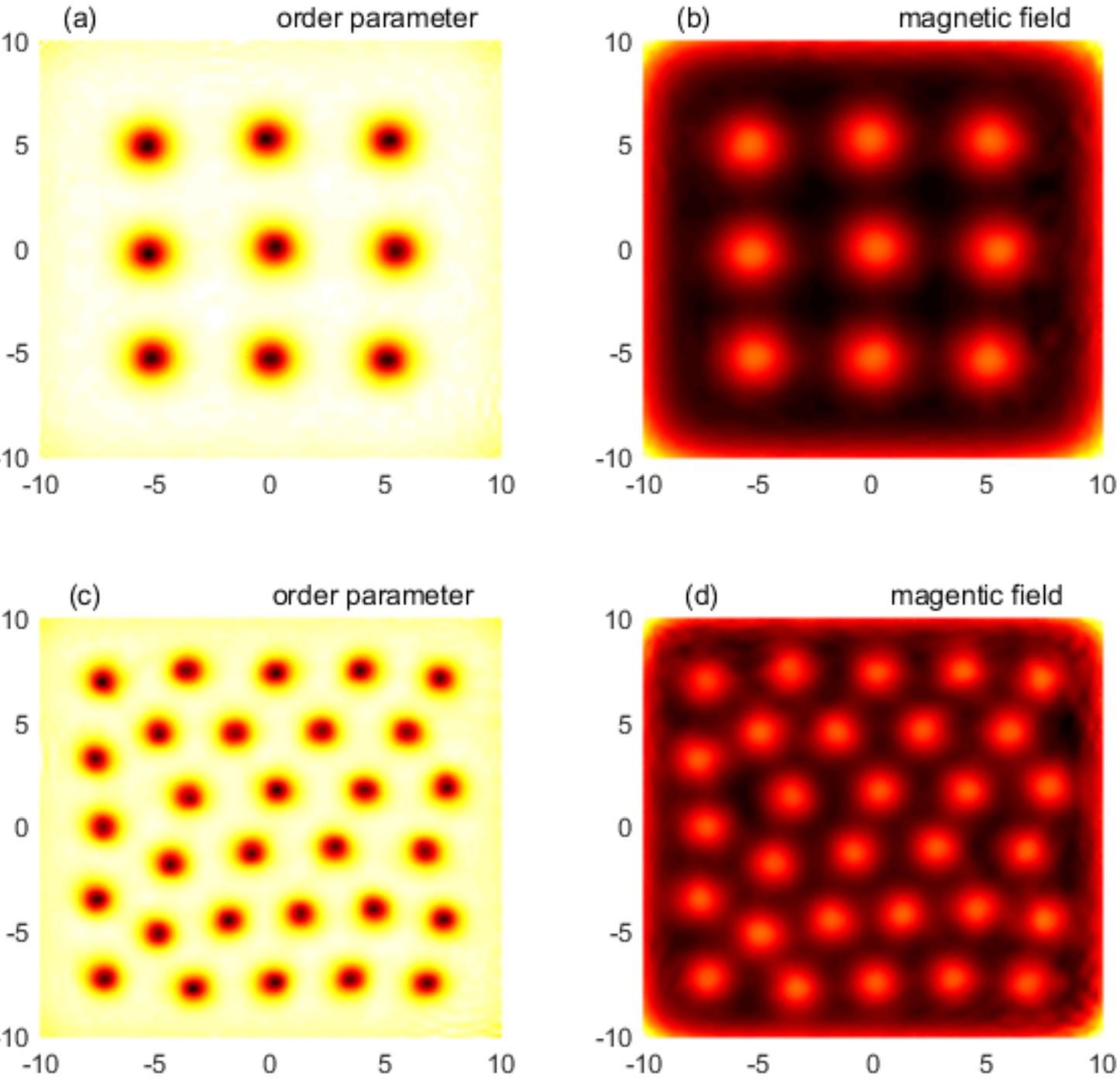}
\caption{Two representative vortex lattice solutions: (a-b) $T = 0.9 \, T_c, \; B_0 = 0.8$,
and (c-d) $T = 0.82 \, T_c, \; B_0 = 1.7$.
} \label{fig_config2}
\end{figure}

However, different things happen when the vortex number is increased by a larger magnetic field. Two typical examples are given in Fig.~\ref{fig_config2}. For the case with $T = 0.9 \, T_c, B_0 = 0.8$, there are 9 vortices forming a square pattern. In this case, the distance between vortices, $a_\square \approx 5.02$, close to the value from~(\ref{atrig}) $a_\square \approx 5.05$, 
is not small enough and the finite size of system prohibits the formation of a hexagon pattern. For the case $T = 0.82 \, T_c, B_0 = 1.7$, the vortex number is 30, the distance between two nearest vortices computed by~(\ref{atrig}) $a_{\triangle} \approx 3.43$ 
is also close to the numerical simulation $a_\triangle \approx 3.214$. The finite size effect is moderated, thus the hexagonal pattern is favored. However, the array does not admit a perfect lattice configuration, which should be a consequence of boundary effects since the vortices are close to the boundary. Keep increasing the magnetic field to the $B_{c2}$, many superconducting areas undergo a phase transition to normal state, leaving a superconducting island with vortices crowded together without a well defined nearest vortex distance.

\begin{figure}
\centering
\includegraphics[trim=3cm 12cm 7.5cm 11cm, clip=true, scale=0.6, angle=0]{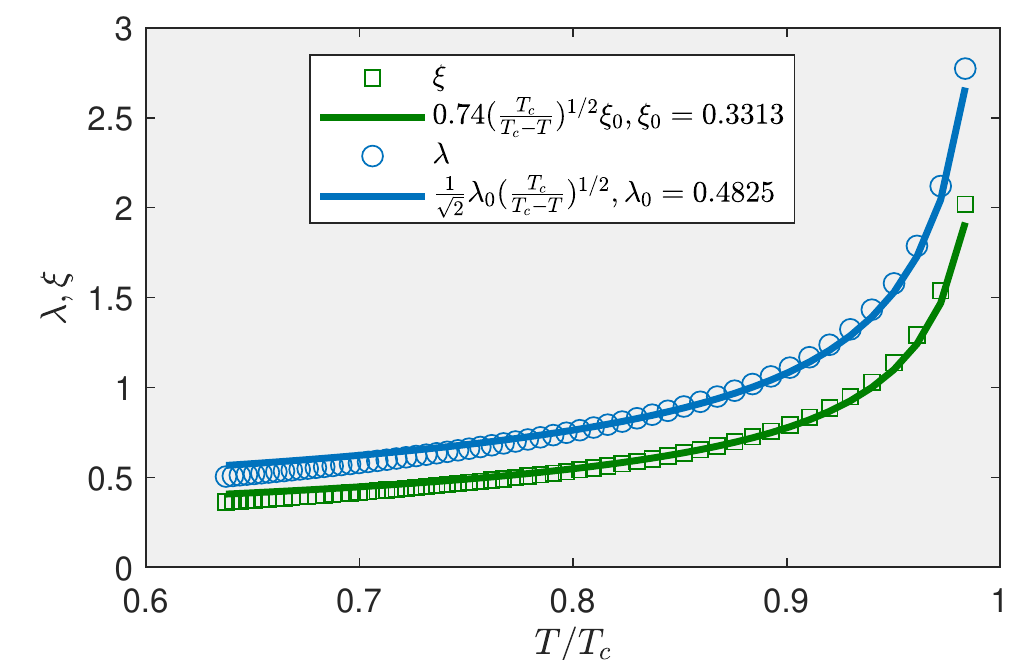}
\caption{The temperature dependence of $\xi, \lambda$ with fitting formulae in~(\ref{xilambda}).} \label{fig_xilambda}
\end{figure}

One can also clearly observe that the vortex size enlarges when the temperature is increasing. More precisely, from the size of the vortex, and similarly of the magnetic field profile, we can read out the dependence of $\lambda$ and $\xi$ with respect to $T/T_c$ as shown in Fig.~\ref{fig_xilambda}. Their behaviors are, almost independent on $B_0$, consistent with the results by GL theory~\cite{Tinkham} for $T \approx T_c$
\begin{equation} \label{xilambda}
\xi \sim 0.74 \xi_0 (1 - T/T_c)^{-1/2}, \quad \lambda \sim \frac{\lambda_0}{\sqrt2} (1 - T/T_c)^{-1/2},
\end{equation}
with $\xi_0 \sim 0.3313$ and $\lambda_0 \sim 0.4825$. Thus the GL parameter generally is $\kappa \sim 1.3916$. It is worth to note that these formulas can fit the data for a broad range of temperature away from $T_c$, for example with about 3\% variation to the value at the point $T = 0.95 T_c$.

\begin{figure}
\centering
\includegraphics[trim=6cm 12.7cm 6cm 13.5cm, clip=true, scale=1.1, angle=0]{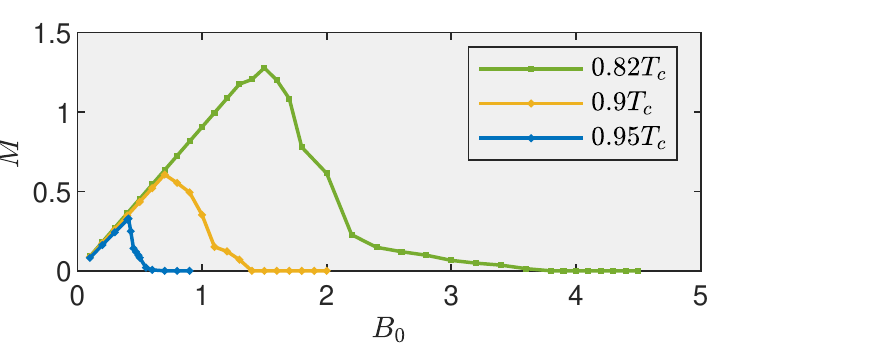}
\caption{The magnetisation $M$ defined in Eq.~(\ref{MofB}) for three different temperatures $T = 0.82 \, T_c, 0.9 \, T_c $ and $0.95 \, T_c$, from the curves we can read the two critical fields where the magnetisation reaches its maximal value and reduces to zero respectively.} \label{fig_M}
\end{figure}

\textit{Magnetisation} ---
To distinguish the Meissner phase and the vortex lattice phase, we can compute the  magnetisation $M$, which is defined as
\begin{equation} \label{MofB}
M(B_0) = B_0 - \left\langle B(x, y) \right\rangle,
\end{equation}
where the $B_0$ is the value of the external magnetic field applied at the initial time, $B(x, y)$ is the magnetic field distribution in the final equilibrium state. When the added external field increases from zero to $B_{c1}$, the magnetic field is completely excluded then there is no magnetic field inside the sample, i.e. $M = B_0$. While above $B_{c2}$, the superconductivity is already completely destroyed, therefore $B(x, y) = B_0$ and $M$ should be zero. In the mixed state $B_{c1} < B_0 < B_{c2}$, the magnetization $M$ will decrease from $B_{c1}$ to zero gradually. In Fig.~\ref{fig_M} we show the magnetization versus $B_0$ for three different temperatures.

\begin{figure}
\centering
\includegraphics[trim=4cm 11cm 10cm 11.5cm, clip=true, scale=0.8, angle=0]{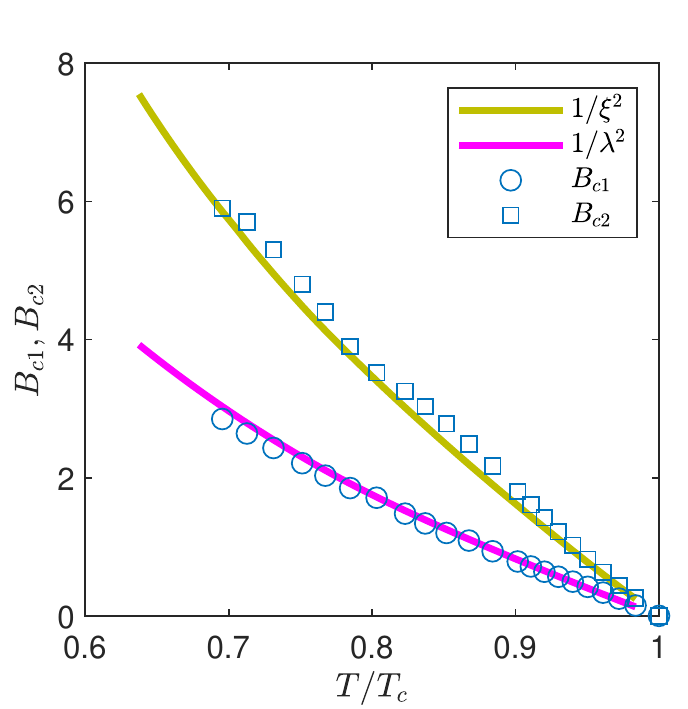}
\caption{The phase diagram of holographic superconductor under external magnetic field.} \label{fig_phase}
\end{figure}

\textit{Phase Diagram} ---
From the magnetization curves we are able to read the two critical magnetic fields $B_{c1}$ and $B_{c2}$ then to obtain the phase diagram which is shown in Fig.~\ref{fig_phase}. The upper critical field can be naively estimated. As the magnetic field increases, more vortices enter and the lattice becomes more compressed. At a certain point, the cores of vortices overlap and no superconducting path is left for a transport current. Indeed, the second critical field in the holographic superconductor model is
\begin{equation}
B_{c2} \approx 16.64 \, (1 - T/T_c),
\end{equation}
which, according to the result~(\ref{xilambda}), confirms the relation derived by the GL theory~\cite{Blatter:1994zz, Tinkham}
\begin{equation}
B_{c2} = \frac{\Phi_0}{2 \pi \xi^2} = \xi^{-2},
\end{equation}
where $\pi \xi^2$ is the size of the Abrikosov unit cell. Moreover, the lower critical magnetic field $B_{c1}$ is close to the intuitive estimation for the moment when the first vortex was created
\begin{equation}
B_{c1} \approx \frac{\Phi_0}{2 \pi \lambda^2} = \lambda^{-2},
\end{equation}
where $\pi \lambda^2$ is the magnetic field penetrated area.

\textit{Summary} ---
Before the advent of AdS/CFT correspondence and the holographic superconductor model defined in the AdS/CFT correspondence framework, the study of the Abrikosov lattice formation dynamics was mainly based on the Gindzburg-Landau theory. We find that the holographic superconductor model offers another approach that the dynamics of magnetic quantum fluxes in spatial two dimensions can be captured by solving the highly nonlinear coupled PDEs in the bulk geometry. All the results agree with GL theory and the experimental observations. There are many other issues needed to be studied, for example, a detailed study of vortex matter near $B_{c2}$ may find the vortex liquid state, extending the model to AdS$_5$ will enable us to study the vortex lines dynamics in high temperature superconductors. Also we focus on the final equilibrium vortex lattice configuration in the 
present work, the study of detailed vortex formation dynamics  is in progress. 

\textit{Acknowledgements} ---
This work is supported by the National Natural Science Foundation of China under Grant No. 11675140 (CYX, HBZ), 11975235, 12035016 (YT), and the Ministry of Science and Technology of the R.O.C. under the grants MOST 109-2112-M-008-010, 110-2112-M-008-009 (CMC).


\end{document}